\documentclass{emulateapj}
\usepackage{times}
\pdfoutput = 1

\newcommand{\sig}{$\sigma$}
\newcommand{\hgpc}{$~h^{-3}$Gpc$^3$}
\newcommand{\hmpc}{$~h^{-1}$Mpc}
\newcommand{\muK}{$\mu$K}
\newcommand{\LCDM}{$\Lambda$CDM}
\def\zobov{{\scshape zobov}}
\def\voboz{{\scshape voboz}}

\begin{document}

\title{An Imprint of Super-Structures on the Microwave Background due to
the Integrated Sachs-Wolfe Effect}

\author{Benjamin R. Granett, Mark C. Neyrinck and Istv\'an Szapudi}
\affil{Institute for Astronomy, University of Hawaii, 2680 Woodlawn Drive, Honolulu HI 96822}

\begin{abstract}
We measure hot and cold spots on the microwave background associated
with supercluster and supervoid structures identified in the Sloan
Digital Sky Survey Luminous Red Galaxy catalog.  The structures give a
compelling visual imprint, with a mean temperature deviation of
$9.6\pm2.2\mu$K, i.e.\ above 4\sig.  We interpret this as a detection
of the late-time Integrated Sachs-Wolfe (ISW) effect, in which cosmic
acceleration from dark energy causes gravitational potentials to
decay, heating or cooling photons passing through density crests or
troughs.  In a flat universe, the linear ISW effect is a direct signal
of dark energy.
\end{abstract}

\keywords{cosmic microwave background --- cosmology: observations ---
large-scale structure of universe --- methods: statistical}

\section{Introduction}
The cosmic microwave background (CMB) is a snapshot of the early
universe; however, the light we observe has been processed by
large-scale structure at low redshift, in part through the late-time
integrated Sachs-Wolfe (ISW) effect \citep{SachsWolfe}.  As photons
travel through time-varying gravitational potentials, they are
slightly heated or cooled.  In a universe dominated by dark energy,
the gravitational potential decays with time even in linear theory,
heating photons traveling through crests and cooling photons in
troughs of large-scale matter density fluctuations.  Hereafter, `ISW'
refers to the full nonlinear late-time ISW effect, also known as the
Rees-Sciama effect \citep{ReesSciama}.

The ISW effect from dark energy can be detected with the
cross-correlation function between the projected galaxy density and
microwave background temperature over the sky \citep{Crittenden}.
Measurements from individual galaxy surveys detect the effect with
signal-to-noise no higher than 3
\citep{Scranton,Boughn,Afsh2MASS,Pad,Racc}.  Recently, various groups
have combined multiple datasets to arrive at a detection as high as
4.5\sig, though error estimation with correlated galaxy datasets
complicates the physical interpretation \citep{Ho,Giannantonio}.
Additionally, studies using wavelet analyses have suggested that the
signal can be localized to particular regions on the sky that depend
on both the CMB and the galaxy density \citep{mcewen}.

The ISW signal peaks at spherical multipole $\ell\sim 20$ at $z=0.5$ in
the galaxy-CMB cross-power spectrum
$\ell(\ell+l)C_\ell$\ \citep[e.g.][]{Pad}.  This corresponds to
structures with angular radius $\sim 4$\degr, or $\sim 100$\hmpc.  We
call these large structures `supervoids' and `superclusters,' but they
may be thought of as gentle hills and valleys in the linear density
field.  In a \LCDM\ universe, the ISW signal from these broad, linear
over- or under-dense structures is expected to dominate over
smaller-scale fluctuations in the density.

In this study, we identified a sample of supervoids and superclusters
in a galaxy survey that could potentially produce measurable ISW
signals.  We analyze these structures by stacking cutouts of the CMB
centered on their projected locations.  \citet{DaleHarald} recently
used a similar method on much smaller spatial scales, stacking WMAP
data behind clusters to detect the frequency-dependent thermal
Sunyaev-Zeldovich \citep{SZ} effect.  We show that our structures are,
on average, associated with a significant temperature imprint on the
CMB.  This is arguably the first visually compelling detection of the
ISW effect.  In our conclusions, we discuss the application of this
work to dark energy and analysis of secondary CMB anisotropies.

\section{Data and Methodology}
We used a sample of 1.1 million Luminous Red Galaxies (LRGs) from the
Sloan Digital Sky Survey (SDSS) \citep{SDSS} covering 7500 square
degrees about the North Galactic pole.  They span a redshift range of
$0.4 < z < 0.75$, with a median of $\sim0.5$, and inhabit a volume of
about 5\hgpc. LRGs are elliptical galaxies in massive galaxy clusters
representing large dark-matter halos \citep{Blake}, and are thought to
be physically similar objects across their redshift range
\citep{Eisenstein,Wake}.  This makes them excellent, albeit sparse,
tracers of the cosmic matter distribution on scales $\gtrsim 10$ Mpc.
Our sample was selected from photometric data based on the criteria
used in the Mega-Z LRG catalog over the SDSS Data Release 4
footprint \citep{Collister}.  We remove sources classified as stars in
the SDSS catalog, but do not use the star/galaxy classifier in the
MegaZ catalog.  Contamination by stars is estimated by
\citet{Collister} to be 5\%.  We extended the catalog with the
additional area from Data Release 6.  Redshifts for the new area were
estimated by a nearest neighbor match with the $ugriz$ photometry.  We
estimate that this procedure smooths the redshift distribution by
$\sigma_z = .003$ and has little effect on the overall redshift
uncertainty, which is at the $\sigma_z=.05$ level.

\begin{figure*}
  \begin{minipage}{175mm}
    \begin{center}
      \includegraphics[scale=0.83]{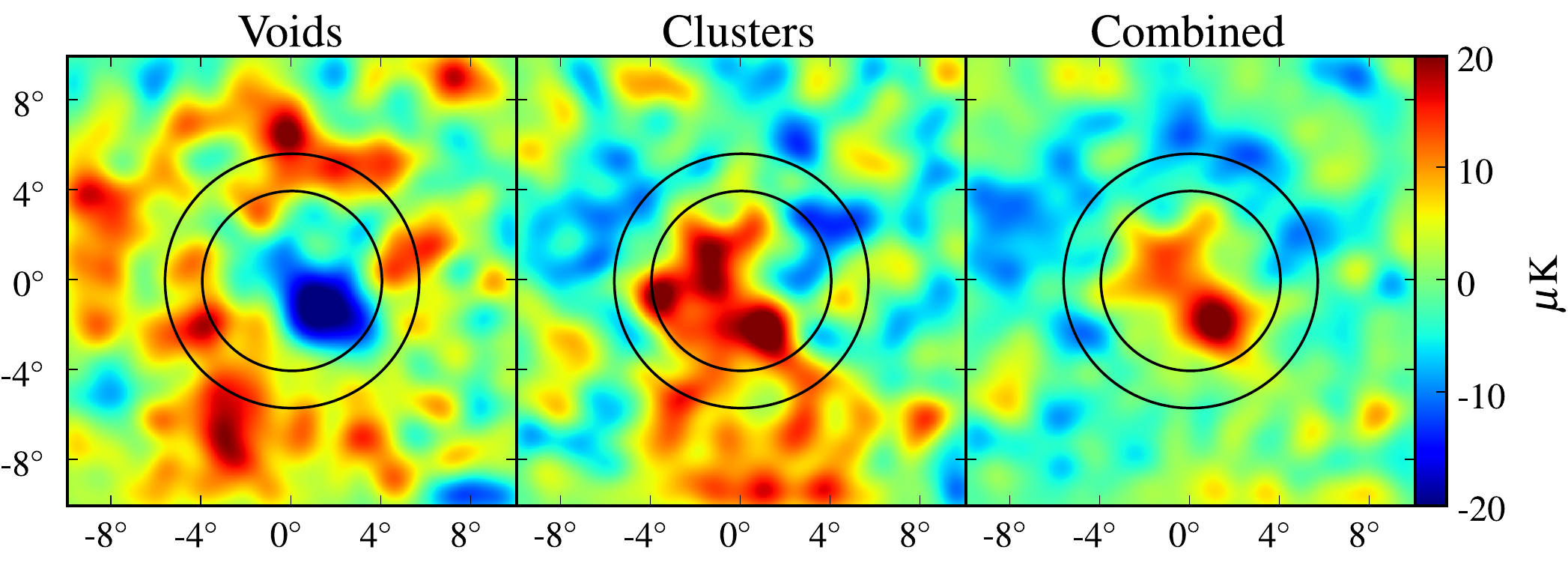}
    \end{center}  
    \caption{\small Stacked regions on the CMB corresponding to supervoid and
      supercluster structures identified in the SDSS LRG catalog.  We
      averaged CMB cut-outs around 50 supervoids (left) and 50 superclusters
      (center), and the combined sample (right).  The cut-outs are rotated,
      to align each structure's major axis with the vertical direction.  Our
      statistical analysis uses the raw images, but for this figure we
      smooth them with a Gaussian kernel with FWHM 1.4\degr.  Hot and cold
      spots appear in the cluster and void stacks, respectively, with a
      characteristic radius of 4\degr, corresponding to spatial scales of
      100\hmpc.  The inner circle (4\degr\ radius) and equal-area outer ring
      mark the extent of the compensated filter used in our analysis.  Given
      the uncertainty in void and cluster orientations, small-scale features
      should be interpreted cautiously.\label{fig:stack}}
  \end{minipage}
\end{figure*}

The CMB temperature map we used is an inverse-variance weighted
combination of the {\it Wilkinson Microwave Anisotropy Probe} (WMAP) 5-year
Q, V and W frequency maps \citep{Hinshaw}, with the foreground
galactic-emission maps subtracted from each.  Regions within the
extended temperature analysis mask (KQ75), which is a conservative
Galactic and point source mask, are left out of the analysis.  The
maps are pixelized in Healpix format \citep{Healpix} at 7 arcminute
resolution, which oversamples the 30-arcminute full-width, half-max
beam.  In excellent agreement with previous results
\citep{Giannantonio}, we measured a cross-correlation amplitude
between our two data sets on 1\degr\ scales of 0.7\muK.

To find supervoids in the galaxy sample, we used the parameter-free,
publicly available \zobov\ \citep[ZOnes Bordering On
Voidness;][]{zobov} algorithm.  For each galaxy, \zobov\ estimates the
density and set of neighbors using the parameter-free Voronoi
tessellation \citep{okabe,vdws}.  Then, around each density minimum,
\zobov\ finds density depressions, i.e.\ voids.  We used \voboz\
\citep{voboz} to detect clusters, the same algorithm applied to the
inverse of the density.

In 2D, if density were represented as height, the density depressions
\zobov\ finds would correspond to catchment basins
\citep[e.g.][]{pwj}. Large voids can include multiple depressions,
joined together to form a most-probable extent. This requires judging
the significance of a depression; for this, we use its density
contrast, comparing against density contrasts of voids from a uniform
Poisson point sample.  Most of the voids and clusters in our catalog
consist of single depressions.

We estimated the density of the galaxy sample in 3D, converting
redshift to distance according to WMAP5 \citep{komatsu} cosmological
parameters. To correct for the variable selection function, we
normalized the galaxy densities to have the same mean in 100 equally
spaced distance bins.  This also removes almost all dependence on the
redshift-distance mapping that the galaxy densities might have.  We
took many steps to ensure that survey boundaries and holes did not
affect the structures we detected.  We put a 1\degr\ buffer of
galaxies (sampled at thrice the mean density) around the survey
footprint, and put buffer galaxies with maximum separation
1\degr\ from each other in front of and behind the dataset.  Any real
galaxies with Voronoi neighbors within a buffer were not used to find
structures.  We handled survey holes (caused by bright stars,
etc.)\ by filling them with random fake galaxies at the mean density.
The hole galaxies comprise about 1/300 of the galaxies used to find
voids and clusters.  From the final cluster and void lists, we
discarded any structures that overlapped LRG survey holes by
$\ge$10\%, that were $\le2.5$\degr\ (the stripe width) from the
footprint boundary, that were centered on a WMAP point source, or that
otherwise fell outside the boundaries of the WMAP mask.

We found 631 voids and 2836 clusters above a 2\sig~significance level,
evaluated by comparing their density contrasts to those of voids and
clusters in a uniform Poisson point sample.  There are so many
structures because of the high sensitivity of the Voronoi
tessellation.  Most of them are spurious, arising from discreteness
noise.  We used only the highest-density-contrast structures in our
analysis; we discuss the size of our sample below.

We defined the centers of structures by averaging the positions of
member galaxies, weighting by the Voronoi volume in the case of voids.
The mean radius of voids, defined as the average distance of member
galaxies from the center, was 2.0\degr; for clusters, the mean radius
was 0.5\degr.  The average maximum distance between void galaxies and
centers was 4.0\degr; for clusters, it was 1.1\degr.  For each
structure, an orientation and ellipticity is measured using the
moments of the member galaxies, though it is not expected that this
morphological information is significant, given the galaxy sparseness.

\section{Imprints on the CMB}

\begin{figure}
\epsscale{1} \plotone{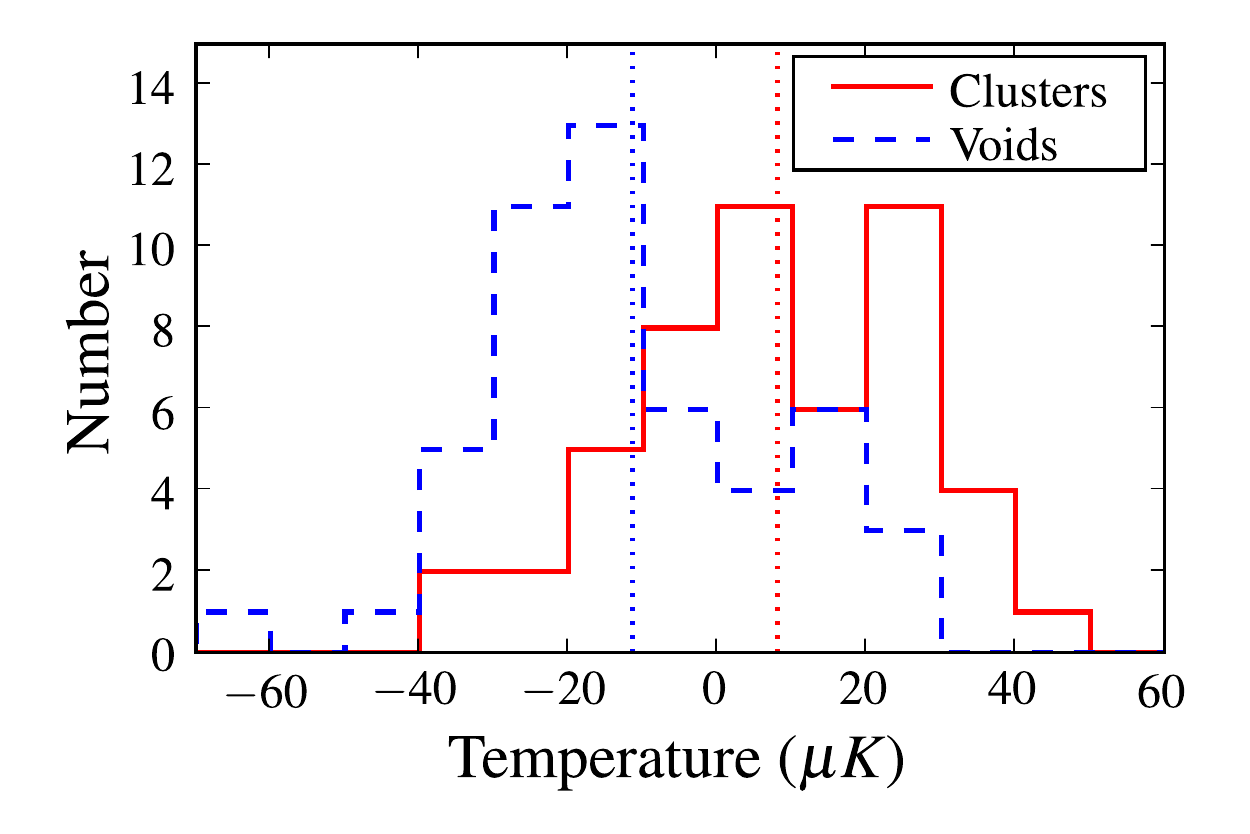}
\caption{\small Histograms of the signals of the 50 highest-significance
  supervoids and superclusters used for our detection, measured in our
  compensated filter.  The vertical dotted lines are the means of each
  distribution, at -11.3$\pm3.1$\muK (voids) and 7.9$\pm3.1$\muK
  (clusters).
\label{fig:temp_hist}
}
\end{figure}

Figure \ref{fig:stack} shows a stack image built by averaging the
regions on the CMB surrounding each object.  The CMB stack
corresponding to supervoids shows a cold spot of -11.3\muK\ with
3.7\sig\ significance, while that corresponding to superclusters shows
a hot spot of 7.9\muK\ with 2.6\sig~significance, assessed in the same
way as for the combined signal, described below.  Figure
\ref{fig:temp_hist} shows a histogram of the signals from each void
and cluster.

To assess the significance of our detection, we averaged the negative
of the supervoid image with the supercluster image, expecting that the
voids would produce an opposite signal from the clusters.  We used a
top-hat compensated filter to measure the fluctuations, averaging the
mean temperature within 4\degr\ of the center, and then subtracting
the mean temperature in a ring of the same area around it. This filter
is insensitive to CMB fluctuations on scales larger than the object
detected; for an uncompensated filter, these fluctuations would
constitute a significant source of noise.

What is the likelihood that our results are due to random
fluctuations? To estimate that, we performed two sets of 1000 Monte
Carlo simulations. First, we generated random positions of voids and
clusters within the survey and stacked the corresponding areas of the
actual CMB map.  This models the errors given the observed CMB sky and
foreground subtraction, but might not properly account for any
covariance due to the actual configuration of voids and clusters.
Second, we generated model CMB skies smoothed to WMAP resolution and
repeated our analysis on these with the actual void and cluster
configurations observed in the catalogs.  We find that these two
approaches produce identical distributions consistent with Gaussians,
and with standard deviations within 2\% of each other. The hypothesis
that the signal arose from random fluctuations is excluded at the
4.4\sig\ level, a 1:200,000 chance. Our final mean signal with errors
is $9.6\pm2.2$\muK.

We note that the radii of the structures found by \zobov/\voboz\ are
typically less than 4\degr.  One possible reason is that the algorithm
could be conservative in defining edges in the face of significant
discreteness noise.  The detected structures could just be the tips of
larger hills and valleys in the potential.  The stacked signal is also
likely smeared somewhat from noise in determining the structures'
centers.

Our procedure does have two parameters: the number of objects used to
generate the stacked image, and the filter size used to assess the hot
and cold spots' significance.  We used the same number of voids and
clusters for simplicity.  Density-contrast thresholds of 4, 3 and
2\sig\ give 7, 51, and 631 voids, respectively.  With too few
structures, the measurement would be swamped by CMB fluctuations (with
a standard deviation of 22\muK\ in our filter).  With too many,
structures would be introduced that have dubious physicality.  We used
the 50 objects with the highest density contrast (a cut at $\sim
3\sigma$ for voids, and $\sim 3.3\sigma$ for clusters) to roughly
balance these effects.  Stacking 70 voids and clusters gives a signal
of 2.8\sig; with 30, the signal remains above 4\sig.  These results
appear in Table \ref{table:num}.  It is almost certain that some
number of objects would give a higher significance than 50 gives, but
we did not search this parameter space, to simplify the
interpretation.

As mentioned earlier, the ISW signal is expected to peak at a radius
of $\sim 4$\degr\ according to theory; this is also confirmed by the
visual appearance of the images.  Changing the filter radius between
3-5\degr\ results in various detection significances of approximately
4\sig; these results are listed in Table \ref{table:num}.  In a strict
Bayesian sense, even inspecting the image by eye prior to statistical
analysis complicates the interpretation due to {\it a posteriori} bias
issues.  This would be difficult to quantify, but its effect should be
small because the signal was robust in the few cases we checked.

\begin{table}
\begin{center}
\caption{Dependence on number and radius.\label{table:num}}
\begin{tabular}{rrrr}
\tableline
N& Radius & $\Delta T$\muK & $\Delta T / \sigma$\\
\tableline
30 & 4.0\degr & 11.1 & 4.0\\
50 & 4.0\degr & 9.6  & 4.4\\
70 & 4.0\degr & 5.4 & 2.8\\
\tableline
50 & 3.0\degr & 8.4 & 3.4\\
50 & 3.5\degr & 9.3 & 4.0\\
50 & 4.0\degr & 9.6 & 4.4\\
50 & 4.5\degr & 9.2 & 4.4\\
50 & 5.0\degr & 7.8 & 3.8\\
\tableline
\end{tabular}
\end{center}
\end{table}

There are systematic effects from foreground contamination that, in
principle, can mimic the ISW signal.  Dust emission from the Milky Way
is bright at microwave frequencies and is correlated with the dust
extinction correction used in the galaxy catalog.  Extragalactic radio
sources correlated with luminous red galaxies could also potentially
contribute to a false signal at microwave frequencies. To check that
we are not significantly contaminated by microwave sources, we
repeated our analysis on the individual frequency maps using the KQ75
mask as in our combined analysis, but without subtracting the
foreground template maps.  The mean amplitudes of the void signal in
the Q,V and W bands were $-10.6$, $-11.1$ and $-11.1$\ \muK; the mean
amplitudes of the cluster signal were 7.8, 7.9 and 7.7\ \muK; the
error on each of these means is 3.1\ \muK.  These results agree with
our measurement made on the combined map, and demonstrate that there
is no significant frequency dependence of the signal.  Moreover, the
void signal is expected to be less sensitive to foreground
contamination.

\section{Discussion}
We have measured a 4\sig\ temperature deviation on the CMB due to
supervoids and superclusters at $z\sim 0.5$.  The most likely
explanation for this is that we detect the ISW effect.  The linear ISW
effect vanishes in a flat universe without dark energy, and the
higher-order ISW contribution is expected to be significantly lower
than the ISW in \LCDM\ \citep{seljak,tuluie,Crittenden}.  The
consensus in the literature is that detecting the ISW effect signals
the presence of dark energy in a flat Universe.

To estimate the expected effect from ISW in a \LCDM\ universe, we
measured the signal that the Millennium cosmological $N$-body
simulation \citep{Mill} produces.  We ray-traced through the
simulation, summing up the change in potential that a photon would
experience passing through the 500\hmpc\ box in each Cartesian
direction.  In this volume, which is large enough for 1 or 2
supervoids and superclusters, we checked that the linear part of the
ISW signal through the box dominates over higher-order effects.
Centering a $100$\hmpc\ aperture around the maximum ISW signal in the
Millennium volume gives $4.2$\muK, $\sim 2\sigma$ lower than what we
observed in our CMB stack.  Though we only expect these numbers to
agree to within an order of magnitude, we note that most previous ISW
measurements are also somewhat higher than the predicted signal in a
\LCDM\ cosmology \citep{Ho}. While more theoretical studies are needed
to turn our detection into precision constraints on cosmological
parameters, we interpret our image as the ISW effect on the CMB caused
by the decaying of potentials in an accelerating universe with dark
energy.

Previous works used the two-point cross-correlation function of 2D
projected galaxy density maps with the CMB to detect the ISW effect,
reaching a significance of 2 to $2.5\sigma$ for the galaxy sample we
analyzed \citep{Ho, Giannantonio}. Several factors likely contribute
to the higher significance of our measurement.  First, we analyze only
superstructures that should be strong ISW sources.  Second, we use 3D
information to identify them.  The 2D projected galaxy density is
typically not extremal at the superstructures' locations; thus, the
cross-correlation function is not especially sensitive to their
contributions.  Third, galaxy autocorrelations directly contribute
to the noise for the cross-correlation function, but not for our
method.

Our detection makes it more plausible that low-redshift supervoids and
superclusters explain anomalies observed on the CMB
\citep{rakic,Rudnick,InoueSilkb,Maturi}.  At low to moderate
significance, these features include a 5\degr\ 70\muK\ cold spot
\citep{Vielva}, the North-South power asymmetry, the low quadrupole
moment, and the alignment of low multipoles \citep{Dragon}.
Additionally, $f_{\rm nl}$, a measure of non-Gaussianity on the CMB,
has been estimated to be positive at low significance in WMAP
\citep{yadav,komatsu}. This indicates a CMB temperature
distribution that is slightly skewed toward low temperatures, as
predicted by a small nonlinear ISW effect that enhances supervoid
signals over superclusters \citep{tomita}.  We indeed find somewhat
stronger cold spots, and although the difference is not statistically
significant, its consistency with the above picture is intriguing.
\\

For supplementary information, including the void and cluster
catalogs, see \citet{Graneypudi}, and
\url{http://ifa.hawaii.edu/cosmowave/supervoids/}.

\acknowledgments
We thank Adrian Pope for useful discussions and help with the SDSS
masks and catalog. We are grateful for support from NASA grant
NNG06GE71G and NSF grant AMS04-0434413.

Funding for the SDSS and SDSS-II has been provided by the Alfred
P. Sloan Foundation, the Participating Institutions, the National
Science Foundation, the U.S. Department of Energy, the National
Aeronautics and Space Administration, the Japanese Monbukagakusho, the
Max Planck Society, and the Higher Education Funding Council for
England. The SDSS Web Site is http://www.sdss.org/.

The SDSS is managed by the Astrophysical Research Consortium for the
Participating Institutions. The Participating Institutions are the
American Museum of Natural History, Astrophysical Institute Potsdam,
University of Basel, University of Cambridge, Case Western Reserve
University, University of Chicago, Drexel University, Fermilab, the
Institute for Advanced Study, the Japan Participation Group, Johns
Hopkins University, the Joint Institute for Nuclear Astrophysics, the
Kavli Institute for Particle Astrophysics and Cosmology, the Korean
Scientist Group, the Chinese Academy of Sciences (LAMOST), Los Alamos
National Laboratory, the Max-Planck-Institute for Astronomy (MPIA),
the Max-Planck-Institute for Astrophysics (MPA), New Mexico State
University, Ohio State University, University of Pittsburgh,
University of Portsmouth, Princeton University, the United States
Naval Observatory, and the University of Washington.

\clearpage

\clearpage

\end{document}